\documentclass[a4paper,10pt]{article}
\usepackage{amsfonts}
\usepackage{amsmath}
\usepackage{amssymb}
\usepackage[dvips]{graphicx}

\DeclareGraphicsExtensions{.bmp,.png,.pdf,.jpg}

\begin{document}

\title{Formation of localized magnetic states in graphene in hollow-site
adsorbed adatoms}
\author{F. Escudero$^{\dag }$, J. S. Ardenghi$^{\dag }$\thanks{%
email:\ jsardenghi@gmail.com, fax number:\ +54-291-4595142}, L. Sourrouille$%
^{\dag }$, P. Jasen$^{\dag }$ and A. Juan$^{\dag }$ \\
IFISUR, Departamento de F\'{\i}sica (UNS-CONICET)\\
Avenida Alem 1253, Bah\'{\i}a Blanca, Buenos Aires, Argentina}
\maketitle

\begin{abstract}
By appplying tight binding model of adatoms in graphene, we study
theoretically the localized aspects of the interaction between transition
metal atoms and graphene. Considering the electron-electron interaction by
adding a Hubbard term in the mean-field approximation, we find the
spin-polarized localized and total density of states. We obtain the coupled
system of equations for the occupation number for each spin in the impurity
and we study the fixed points of the solutions. By comparing the top site
and hollow site adsorption, we show that the anomalous broadening of the
latter allows to obtain magnetization for small values of the Hubbard
parameter. Finally, we model the magnetic boundaries in order to obtain the
range of Fermi energies at which magnetization starts.
\end{abstract}

\section{Introduction}

Graphene is a well-known allotrope of carbon which has become one of the
most fascinating research topics in solid state physics due to the large
number of applications (\cite{novo},\cite{intro1}, \cite{B}). The carbon
atoms bond in a planar sp$_{2}$ configuration forming a honey-comb lattice
made of two interpenetrating triangular sublattices, $A$ and $B$. A special
feature of the graphene band structure is the linear dispersion at the Dirac
points which are dictated by the $\pi $ and $\pi ^{\prime }$ bands that form
conical valleys touching at the two independent high symmetry points at the
corner of the Brillouin zone, the so called valley pseudospin \cite{A}. The
electrons near these symmetry points behave as massless relativistic Dirac
fermions with an effective Dirac-Weyl Hamiltonian \cite{B} and a zero band
gap at the Dirac point. In turn, graphene has been interesting as a 2D model
for carbon-based electronic materials. In the last years, a large number of
experimental and theoretical investigations have been carried out
considering the effects of adatoms and impurities on the band structure and
localized magnetic moments in graphene. These impurities in graphene can be
considered in various types of forms: substitutional, where the site energy
is different from those of carbon atoms, which generates resonances \cite%
{mahan} and as adsorbates, that can be placed on various points in graphene:
six-fold hollow site of a honeycomb lattice, two-fold bridge site of the two
neighboring carbons or top site of a carbon atom \cite{roten}. Theoretical
as well as experimental studies have indicated that substitutional doping of
carbon materials can be used to tailor their physical and/or chemical
properties (\cite{stro}, \cite{jsa}). In particular, theoretical studies on
carbon vacancies in graphene (\cite{leh} and \cite{skr}), adsorbed hydrogen
atoms \cite{sofo}, and several other types of disorder have been done (\cite%
{weh}, \cite{jsa2}, \cite{jsa3} and \cite{jsa4}). From the possible adatoms,
transition metal atoms (TM) have attracted considerable interest in the
fields of hydrogen storage (\cite{yil} and \cite{chandra}), where TM doping
process preserves the structural integrity of carbon nanomaterials, and
therefore, it can be considered as the best alternative for enhancing the
hydrogen storage capacity, in molecular sensing (\cite{kong} and \cite{mota}%
), \ catalysis (\cite{pla} and \cite{li}) and nanoelectronics \cite{jav}.
Moreover, graphene has become a very important material for spintronic
applications given the controllable spin transport \cite{son}, its perfect
spin filtering \cite{karp}, and large lifetime and spin-relaxation lengths
in the order of the micrometre at standard conditions for inyected spins (%
\cite{tom}), due to the very weak spin-orbit coupling in carbon \cite{intro1}%
. Actually, the adsorption of transition metal atoms on graphene is of great
interest since the doping process preserves the integral structure of this
system and also promote the formation of a local magnetic moment due to
states $d$ partially filled \cite{xiao1} which allows to differentiate the
transport properties of the spin channels \cite{bera} with remarkable
implications for the usage of such systems as nanomagnets \cite{liu} and
data storage \cite{xiao1}. In turn, Ru atom interacts strongly with graphene
and locally modifies the charge density in the vicinity of the carbon atoms (%
\cite{verdi} and \cite{luna}). Ru nanoparticles in mesoporous carbon
materials show high catalytic activity in the Fischer-Tropsch synthesis \cite%
{chen}. Magnetic properties of Rh over carbon nanotubes have been done using
ab-initio calculations \cite{luna} showing similar results obtained
experimentally, where magnetic moments in a range of $0.8\mu B$/atom to $%
3\mu B$/atom can be formed with Rh clusters of at least, 60 atoms \cite%
{soltani}. In this sense, the aim of this work is to study the formation of
magnetic moments localized in the adsorbed adatoms in graphene. By using
Green function methods, analytical expressions for the local density of
states (LDOS) on the adatom is obtained and the ocuppation number of each
spin is determined. The magnetic properties of the system are computed using
the Hubbard model for the electron-electron interaction by using a standard
mean-field approximation \cite{anderson}. A set of self-consistent equations
are obtained for the ocuppation number and a detailed study is done for
fixed points of the iteration. Through these results it is possible to
obtain approximate values for the chemical potential and the Hubbard
parameter at which the magnetization arises. This work will be organized as
follow: In section II, the tight-binding model with adatoms is introduced
and the Anderson model in the mean-field approximation is applied. In
section III, the results are shown and a discussion is given and the
principal findings of this paper are highlighted in the conclusion. In
Appendix, the quasiparticle residue and broadening is obtained for hollow
site adsorption.

\section{Theoretical model}

The tight-binding Hamiltonian of graphene for nearest neighbors can be
written as $H_{0}=-t\underset{\left\langle i,j\right\rangle ,\sigma }{%
\overset{}{\sum }}(a_{i,\sigma }^{\dag }b_{j,\sigma }+b_{i,\sigma }^{\dag
}a_{j,\sigma })$, where $a_{i,\sigma }^{\dag }$($a_{i,\sigma }$) creates
(annihilates) an electron on site $\mathbf{r_{i}}$ with spin $\sigma $,
where $\sigma =1,2$ on sublattice $A$ and $b_{i,\sigma }^{\dag }$($%
b_{i,\sigma }$) creates (annihilates)\ an electron on site $\mathbf{r_{i}}$
with spin $\sigma $, on sublattice $B$ and $t=2.8$eV is the nearest neighbor 
$\left\langle i,j\right\rangle $ hopping energy.\footnote{%
Instead of using $\uparrow $ and $\downarrow $ for the spin up and down we
are using the subscript $1$ and $2$ for the sake of simplicity.} By
introducing the Fourier transform of the annihilation and creation operators 
$a_{i,\sigma }=\frac{1}{\sqrt{N}}\underset{\mathbf{k}}{\overset{}{\sum }}e^{i%
\mathbf{k\mathbf{r_{i}}}}a_{\mathbf{k},\sigma }$ and $b_{i,\sigma }=\frac{1}{%
\sqrt{N}}\underset{\mathbf{k}}{\overset{}{\sum }}e^{i\mathbf{k\mathbf{r_{i}}}%
}b_{\mathbf{k},\sigma }$, where $N$ is the number of primitive cells in the
graphene lattice, the Hamiltonian can be written as $H_{0}=-t\underset{%
\mathbf{k,}\sigma }{\overset{}{\sum }}\left[ \phi (\mathbf{k)}a_{\mathbf{k}%
,\sigma }^{\dag }b_{\mathbf{k},\sigma }+\phi ^{\ast }(\mathbf{k)}b_{\mathbf{k%
},\sigma }^{\dag }a_{\mathbf{k},\sigma }\right] $, where $\phi \mathbf{_{%
\mathbf{k}}}=\overset{3}{\underset{i=1}{\sum }}e^{i\mathbf{k\cdot \delta _{i}%
}}$, where the $\delta _{i}$ are the nearest-neighbor bond length, $\mathbf{%
\delta _{1}=-}a\widehat{e}_{x}$, $\mathbf{\delta }_{2}=\frac{a}{2}\widehat{e}%
_{x}+\frac{\sqrt{3}a}{2}\widehat{e}_{y}$ and $\mathbf{\delta _{3}=}\frac{a}{2%
}\widehat{e}_{x}-\frac{\sqrt{3}a}{2}\widehat{e}_{y}$ and $a=1.42\overset{%
\circ }{A}$. To describe isolated adatoms adsorbed onto the host graphene,
we must consider that simulations at room temperature have shown that
adatoms adsorbed on the surface of graphene can be moved from one position
to another, being two minima corresponding energy to the adatom in the
center of the hexagon benzene or adatom located a bridge site. In both
positions, graphene preserves its flatness, with fewer distortions in the
geometry of the C-C bonds near the adatom adsorbed (\cite{man} and \cite%
{ambrusi}). In general, almost all heavy atoms are likely to hybridize at
the hollow site, and most of them hybridize with graphene via $s$, $d$ or $f$
orbitals \cite{neaton}. For simplicity we consider the hollow site, in the
center of the honeycomb hexagon without symmetry breaking, where the adatom
hybridizes with the two sublattices. Considering an adatom in a fixed
position, the hybridization Hamiltonian can be written as%
\begin{equation}
H_{V}=\overset{3}{\underset{i=1,\sigma =1,2}{\sum }}\left[ V_{a,i}a_{\sigma
}^{\dag }(\mathbf{\delta }_{i})+V_{b,i}b_{\sigma }^{\dag }(-\mathbf{\delta }%
_{i})\right] f_{\sigma }+h.c.  \label{4b}
\end{equation}%
The hybridization parameters $V_{x,i}$ are dictated by symmetry only and
represents the orbital involved in the hybridization. In particular $%
V_{a,i}=(-1)^{\gamma }V_{b,i}=V$ where $\gamma =0$ for an $s$-wave orbital
and $\gamma =1$ for a $f$-wave orbital (see \cite{ucho}). By applying the
Fourier transform, last Hamiltonian can be written as $H_{V}=\overset{2}{%
\underset{\sigma =1,2}{\sum }}\left[ V_{a,\mathbf{k}}a_{\mathbf{k}\sigma
}^{\dag }+V_{b,\mathbf{k}}b_{\mathbf{k}\sigma }^{\dag }\right] f_{\sigma
}+h.c.$,where we have used that (see \cite{ucho3})%
\begin{equation}
V_{a,\mathbf{k}}=\overset{3}{\underset{i=1}{\sum }}V_{a,i}e^{i\mathbf{%
k\delta _{i}}}=V\phi _{\mathbf{k}}^{\ast }\text{ \ \ \ \ \ \ \ \ \ \ \ }V_{b,%
\mathbf{k}}=\overset{3}{\underset{i=1}{\sum }}V_{b,i}e^{i\mathbf{k\delta _{i}%
}}=(-1)^{\gamma }V\phi _{\mathbf{k}}  \label{4d}
\end{equation}%
where the sum in $i$ represent summation over the hybridization amplitudes
of the adatom with the nearest neighbor carbon atoms on a given sublattice.
Finally, we can add the interaction between impurities and the Hubbard term
through a Hamiltonian $H_{F}=\epsilon _{0}f_{\sigma }^{\dag }f_{\sigma
}+Un_{1}n_{2}$, where $\epsilon _{0}$ is the single electron energy at the
impurity, $n_{\sigma }=$ $f_{\sigma }^{\dag }f_{\sigma }$ is the occupation
number operator for the impurity and $U$ is the strength of the electron
correlations in the inner shell states of impurities. By adopting the mean
field approximation (\cite{ander}), we can decompose the electronic
correlations at the impurities $Un_{1}n_{2}\sim U\overset{}{\underset{\sigma 
}{\sum }}\left\langle n_{\sigma }\right\rangle f_{\sigma }^{\dag }f_{\sigma
}-U\left\langle n_{1}\right\rangle \left\langle n_{2}\right\rangle $, such
that the impurities Hamiltonian can be rewritten as $H_{F}=\epsilon _{\sigma
}f_{\sigma }^{\dag }f_{\sigma }$, where $\epsilon _{1}=\epsilon
_{0}+U\left\langle n_{2}\right\rangle $ and $\epsilon _{2}=\epsilon
_{0}+U\left\langle n_{1}\right\rangle $. By introducing a new set of
operators $c_{\sigma ,\mathbf{k}}^{(\pm )}=\frac{1}{\sqrt{2}}(b_{\sigma ,%
\mathbf{k}}\pm \frac{\phi _{\mathbf{k}}^{\ast }}{\left\vert \phi _{\mathbf{k}%
}\right\vert }a_{\sigma ,\mathbf{k}})$ the non-interacting Hamiltonian $%
H_{0} $ can be diagonalized in the new basis. In this case, $H_{0}$ and $%
H_{V}$ reads%
\begin{equation}
H_{0}=\underset{\mathbf{k,}\sigma }{\overset{}{\sum }}\left( \epsilon _{%
\mathbf{k}}c_{\sigma ,\mathbf{k}}^{(+)\dag }c_{\sigma ,\mathbf{k}%
}^{(+)}-\epsilon _{\mathbf{k}}c_{\sigma ,\mathbf{k}}^{(-)\dag }c_{\sigma ,%
\mathbf{k}}^{(-)}\right)  \label{6}
\end{equation}%
and%
\begin{equation}
H_{V}=\overset{2}{\underset{\alpha =\pm 1,\mathbf{k,}\sigma }{\sum }}\Theta
_{\mathbf{k\alpha }}c_{\alpha \sigma ,\mathbf{k}}^{\dag }f_{\sigma }+h.c.
\label{6.01}
\end{equation}%
where%
\begin{equation}
\Theta _{\mathbf{k\alpha }}=\frac{V}{\sqrt{2}}\left( \phi _{\mathbf{k}%
}+\alpha (-1)^{\gamma }\frac{\phi _{\mathbf{k}}^{\ast }\phi _{\mathbf{k}%
}^{\ast }}{\left\vert \phi _{\mathbf{k}}\right\vert }\right)  \label{6.02}
\end{equation}%
which is the generalization of eq.(5) of \cite{ucho}. The Hamiltonian $H_{V}$
implies that each impurity hybridize with the valence and conduction band
with hybridization parameter $\Theta _{\mathbf{k\alpha }}$. In order to
study the localized magnetic states, the occupation number of the electron
spins $\sigma $ at the impurities must be computed. The number of states
below the Fermi level $\mu $ are completely occupied and the occupation
number at the impurity reads%
\begin{equation}
n_{\sigma }=\int_{-D}^{\mu }\rho _{\sigma }(\omega )d\omega  \label{6.03}
\end{equation}%
where $\rho _{\sigma }(E)=-\frac{1}{\pi }\Im {G}_{\sigma }(E)$ is the local
density of states at the impurity, where $G_{\sigma }$ is the Green function
at the impurity level and $D\sim 7$eV is the bandwidth. By solving the
coupled algebraic system for the Green function matrix elements, $G_{\sigma
} $ reads%
\begin{equation}
G_{\sigma }=\frac{1}{z-\epsilon _{\sigma }-\Delta }  \label{6.04}
\end{equation}%
where $z=\omega +i0^{+}$ and%
\begin{equation}
\Delta =\overset{}{\underset{k}{\sum }}\left( \frac{\left\vert \Theta _{%
\mathbf{k+}}\right\vert ^{2}}{z-\epsilon _{\mathbf{k}}}+\frac{\left\vert
\Theta _{\mathbf{k-}}\right\vert ^{2}}{z+\epsilon _{\mathbf{k}}}\right)
\label{6.05}
\end{equation}%
The local density of states in the magnetic impurities read%
\begin{equation}
\rho _{\sigma }(\epsilon )=\frac{\Im \Delta }{(Z^{-1}(\epsilon )\epsilon
-\epsilon _{\sigma })^{2}+\Im ^{2}\Delta }  \label{6.07a}
\end{equation}%
where $Z^{-1}(\epsilon )=1+\frac{{\Re }\Delta }{\epsilon }$ is the
quasiparticle residue. 
\begin{figure}[tbp]
\centering\includegraphics[width=120mm,height=45mm]{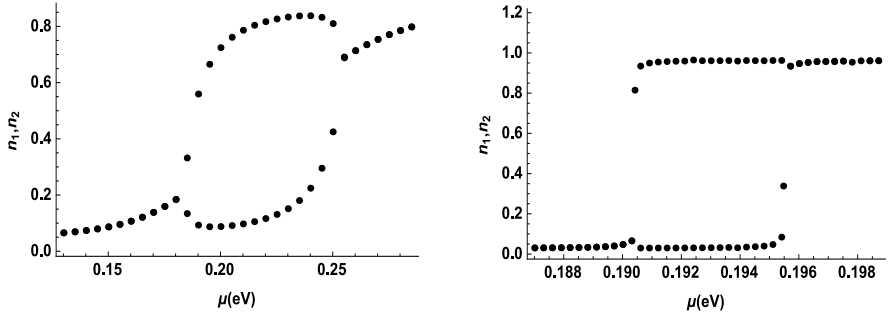}
\caption{Occupation number $n_{1}$ and $n_{2}$ as a function of $\protect\mu 
$ for the adatom adsorbed in a top site (right figure) and in a hollow site
(left figure). $\protect\epsilon _{0}/D=0.029$, $V/D=0.14$. $U=0.1$eV
(right)\ and $U=0.006$eV (left).}
\label{mag1}
\end{figure}
For adsorption on top and hollow sites, the quasiparticle residue and the
hibridization can be written as%
\begin{equation}
Z_{T}^{-1}=1+\pi \xi _{T}\ln (\frac{\left\vert D^{2}-\omega ^{2}\right\vert 
}{\omega ^{2}})\text{ \ \ \ \ \ }\Im \Delta _{T}=\xi _{T}\left\vert \omega
\right\vert  \label{7.0}
\end{equation}%
\begin{equation}
Z_{H}^{-1}=1+\pi \xi _{H}\left[ D^{2}+\omega ^{2}\ln (\frac{\omega ^{2}}{%
\omega ^{2}-D^{2}})\right] \text{ \ \ \ \ \ }\Im \Delta _{H}=\xi
_{H}\left\vert \omega \right\vert ^{3}  \label{7.1}
\end{equation}%
where $\xi _{T}=\pi \frac{V^{2}}{D^{2}}$ and $\xi _{H}=\frac{\pi V^{2}}{%
t^{2}D^{2}}$ for top and hollow site respectively. In general, the formation
of a magnetic moment is determined by the occupation of the two spin states
at the impurity, whenever $n_{1}\neq n_{2}$. The determination $n_{1}$ and $%
n_{2}$ demands to solve the self-consistent calculation of the density of
states of eq.(\ref{6.07a}) at the impurity level, which incorporates the
broadening of the impurity level (eq.\ref{6.05}) due to hybridization with
the bath of electrons in graphene (see \cite{ucho} and \cite{ucho2}). For
the sake of simplicity, in figure \ref{mag1}, the occupation number can be
computed for the top and hollow site adsorption for $U=0.1$eV (top site) and 
$U=0.006$eV and it can be seen that a magnetic moment appears for negligible 
$U$ in a narrow range of $\mu $.

\section{Results and discussions}

\begin{figure}[tbp]
\centering\includegraphics[width=85mm,height=60mm]{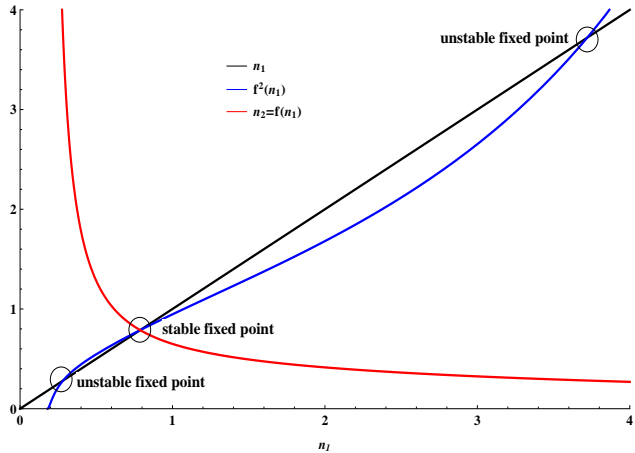}
\caption{$n_{1T}$ (black), $f_{T}(n_{1})$ (red) and $f_{T}^{2}(n_{1})$
(blue) as a function of $n_{1}$ for different values of $\protect\mu $ and $%
U=0.1$eV (top site). }
\label{fixedu1}
\end{figure}

\begin{figure}[tbp]
\centering\includegraphics[width=115mm,height=80mm]{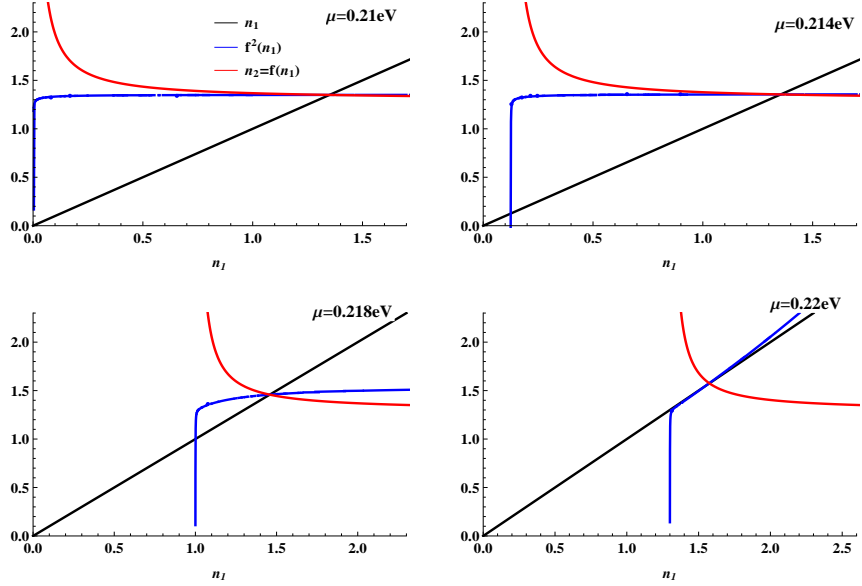}
\caption{$n_{1}$ (black), $f_{H}(n_{1})$ (red) and $f_{H}^{2}(n_{1})$ (blue)
as a function of $n_{1}$ for different values of $\protect\mu $ and $U=0.008$%
eV (hollow site). }
\label{fixedu3}
\end{figure}

\begin{figure}[tbp]
\centering\includegraphics[width=115mm,height=45mm]{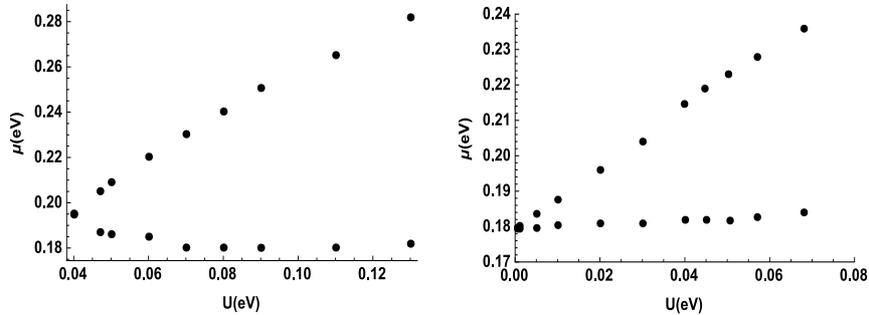}
\caption{Magnetic boundaries for adatom adsorbed in top site (right) and
adsorbed in hollow site\ (left) for $\protect\epsilon _{0}/D=0.029$ and $%
V/D=0.14$. }
\label{fase}
\end{figure}

For typical values, $V/D\sim 0.1$, which implies that $\xi _{T}<<1$ and $\xi
_{H}<<1$. Then it is possible to expand the density of states up to linear
order in $\xi _{T}$ or $\xi _{H}$. In this case, the quasiparticle residue
for both sites is $Z_{T}^{-1}=Z_{H}^{-1}=1$ and the ocuppation number for
the top site adsorption reads%
\begin{gather}
\frac{1}{\xi _{T}}n_{1/2}^{T}=\int_{-D}^{\mu }\frac{\left\vert \omega
\right\vert }{(\omega -\epsilon _{0}-Un_{2/1}^{T})^{2}}d\omega =  \label{r1}
\\
=-2+\epsilon _{0}+Un_{2/1}^{T}S(n_{2/1}^{T})+\ln (R(n_{2/1}^{T}))  \notag
\end{gather}%
and for the hollow site adsorption%
\begin{gather}
\frac{1}{\xi _{H}}n_{1/2}^{H}=\int_{-D}^{\mu }\frac{\left\vert \omega
\right\vert ^{3}}{(\omega -\epsilon _{0}-Un_{2/1}^{H})^{2}}d\omega =
\label{r2} \\
=\frac{D^{2}+\mu ^{2}}{2}+2(\epsilon _{0}+Un_{2/1}^{H})(\mu -D)+  \notag \\
\lbrack \epsilon _{0}+Un_{2/1}^{H}]^{3}S(n_{2/1}^{H})+(\epsilon
_{0}+Un_{2/1}^{H})^{2}\left( 3\ln [R(n_{2/1}^{H})]-2\right)  \notag
\end{gather}%
\begin{figure}[tbp]
\centering\includegraphics[width=110mm,height=75mm]{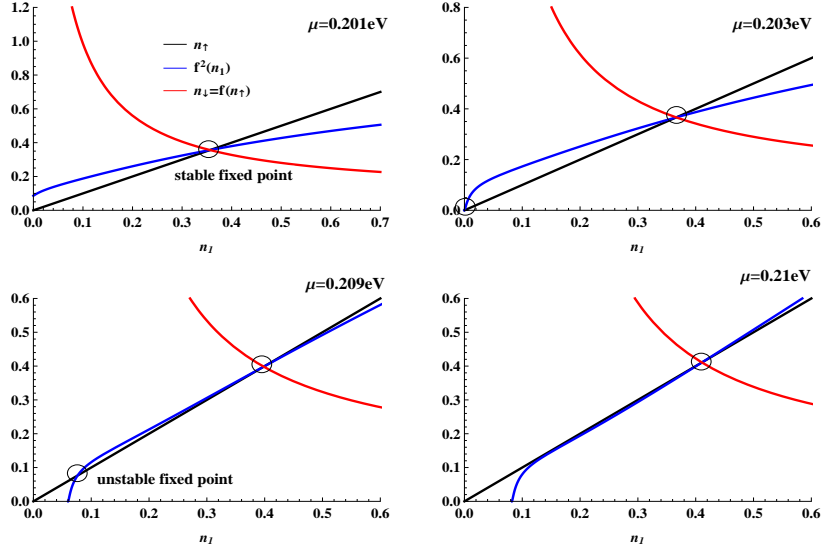}
\caption{$n_{1}$ (black), $f_{T}(n_{1})$ (red) and $f_{T}^{2}(n_{1})$ (blue)
as a function of $n_{1}$ for different values of $\protect\mu $. }
\label{fixed1}
\end{figure}
where%
\begin{equation}
S(n_{2/1})=\frac{D-\mu +2\epsilon _{0}+2Un_{2/1}}{(D+\epsilon
_{0}+Un_{2/1})(\epsilon _{0}+Un_{2/1}-\mu )}  \label{r3}
\end{equation}%
and%
\begin{equation}
R(n_{2/1})=\frac{(D+\epsilon _{0}+Un_{2/1})(\epsilon _{0}+Un_{2/1}-\mu )}{%
(\epsilon _{0}+Un_{2/1})^{2}}  \label{r4}
\end{equation}%
In order to show how the magnetism arise, it can be noted that $%
n_{1}=f(n_{2})$ and $n_{2}=f(n_{1})$ which implies that $%
n_{1}=f^{2}(n_{1})=f(f(n_{1}))$ which it can be solved for $n_{1}$, where $f$
is the function obtained in eq.(\ref{r1}) for both sites adsorption. In
figure \ref{fixed1}, $n_{1}$, $f(n_{1})$ and $f_{T}^{2}(n_{1})~$for $\xi
_{T}=0.06$, $\epsilon _{0}/D=0.029$, $U=0.1$eV and different values of $\mu $
are shown. As it can be seen in all the figures, a stable fixed point can be
obtained for $n_{1}^{s}$ which in turn coincides with $%
n_{1}^{s}=n_{2}^{s}=f_{T}(n_{1}^{s})$. This implies that for this solution
there is no magnetization because $n_{1}^{s}=n_{2}^{s}$. A different
behavior arise when $\mu =0.23$eV, where a different solution appears near
the origin. In this case, the solution obtained is an unstable fixed point
and $n_{2}^{u}=f_{T}(n_{1}^{u})\neq n_{1}^{u}$, which implies that
magnetization should be expected.\footnote{%
The fixed point is unstable because the slope of the function is positive.}
This second solution $n_{1}^{u}$ holds until $\mu =0.3$eV is reached (see
last plot). For $\mu >0.3$eV, only the stable fixed point remains and the
magnetization vanishes. As it can be noted, $n_{2}^{u}$ is larger than $2$
which is expected due to the approximation. In turn, there is a second
unstable fixed point symmetrical to the values $n_{1}^{u}$ and $n_{2}^{u}$
which is shown in figure \ref{fixedu1} for $\mu =0.4$eV, $\epsilon
_{0}/D=0.03$, $V/D=0.14$ and $U=0.08$eV. 
The behavior obtained for the solutions of $n_{1}$ are in concordance with
the results of \cite{ander}, where the unstable fixed points for $n_{1}$ and 
$n_{2}$ are related as follows:\ $n_{2}^{u}=f_{T}(n_{1}^{u})=\overline{n}%
_{1}^{u}$ and $n_{1}^{u}=f_{T}(n_{2}^{u})=\overline{n}_{2}^{u}$. In a
similar way we can proceed with the hollow site adsorption for $\epsilon
_{0}/D=0.03$, $U=0.008$eV, $\xi _{H}=0.028$ and different values of $\mu $
(see figure \ref{fixedu3}), where $n_{1}=f_{H}(n_{2})$ and $%
n_{2}=f_{H}(n_{1})$, where $f_{H}$ is the function defined in eq.(\ref{r2}).
By comparing figure \ref{fixedu1} and figure \ref{fixedu3}, the unstable
solution starts at $n_{1}=0$ when $f(0)\rightarrow \infty $ which \ occurs
for $\mu \sim \epsilon _{0}$. In turn, the slope near the singularity in $%
f^{2}(n)$ is more pronunciated for the hollow site adsorption, even for
small $U$, which allows the system to develop magnetism. 
In order to obtain the critical value of $U$ and $\mu $ at which
magnetization starts, the self-consistent equation for $n_{1}$ and $n_{2}$
can be solved numerically without any approximation for different values of $%
U$. In the anomalous broadening for the hollow site adsorption, the
magnetization appears for $U=0.002$eV, $\epsilon _{0}/D=0.029$ and $V/D=0.14$%
, which implies that the adatom level favors the formation of a local
magnetic moment when $\epsilon _{0}$ is above the Fermi energy, which is
forbidden for ordinary metals. This can be understood by the fact that the
tail of the hybridization decays like $\omega ^{-1}$, which implies a large
broadening of the impurity level density of states that crosses the Fermi
energy even when the bare level energy is above it. In figure \ref{fase},
the boundary between magnetic and non-magnetic states is shown in the $U$
and $\mu $ variables, instead of the common scaling variables $x=\pi
V^{2}/DU $ and $y=(\mu -\epsilon _{0})/U$ used in several works (\cite{gao}%
), for $T$ (left) and $H$ adatom sites (right). From the figure the
similarity between the curves can be seen which implies an universal
behavior for large $U$. In turn, local magnetism is achieved for lower
values of $U$ in $H$ site and the effect of the anomalous broadening is
enhanced. From both figures, the magnetization of the impurity can in
principle be turned on and off, depending only on the gate voltage applied
to graphene. Finally, by considering eq.(\ref{6.03}) for $n_{1}$ and $n_{2}$%
, the magnetization can be written as%
\begin{equation}
n_{1}-n_{2}=\int_{-D}^{\mu }\Im \Delta \left[ \frac{U^{2}\left(
n_{1}^{2}-n_{2}^{2}\right) +2(Z^{-1}\omega -\epsilon _{0})U\left(
n_{2}-n_{1}\right) }{\left[ (Z^{-1}(\epsilon )\omega -\epsilon
_{0}-Un_{2})^{2}+\Im ^{2}\Delta \right] \left[ (Z^{-1}(\epsilon )\omega
-\epsilon _{0}-Un_{1})^{2}+\Im ^{2}\Delta \right] }\right] d\omega
\label{r5}
\end{equation}%
Cancelling $n_{1}-n_{2}$ in both terms of last equation and taking the
expansion at first order in $\xi $, we obtain%
\begin{equation}
1=U\xi \int_{-D}^{\mu }\left\vert \omega \right\vert ^{r}\left[ \frac{%
U(n_{1}+n_{2})-2(\omega -\epsilon _{0})}{(\omega -\epsilon
_{0}-Un_{2})^{2}(\omega -\epsilon _{0}-Un_{1})^{2}}\right] d\omega
\label{r6}
\end{equation}%
where $r=1$ for top site and $r=3$ for hollow site. From last equation we
can obtain two limiting cases for no magnetization, where in both $%
n_{1}=n_{2}=n$. From figure \ref{mag1}, the limiting cases are with $%
n_{1}=n_{2}\sim 0$ and $n_{1}=n_{2}\sim 1$. Replacing in last equation we
obtain two implicit functions of $\mu $ and $U$ which determines the
boundaries where magnetization vanishes%
\begin{equation}
1=-2U\xi \int_{-D}^{\mu }\frac{\left\vert \omega \right\vert ^{r}}{(\omega
-\epsilon _{0})^{3}}d\omega \text{ \ \ \ \ \ \ \ }1=-2U\xi \int_{-D}^{\mu }%
\frac{\left\vert \omega \right\vert ^{r}}{(\omega -\epsilon _{0}-U)^{3}}%
d\omega  \label{r7}
\end{equation}

In particular, for the top site adsorption, $\mu $ reads%
\begin{equation}
\mu _{\pm }^{(n)}=\frac{(\epsilon _{0}+nU)T_{n}\pm \sqrt{T_{n}U(\epsilon
_{0}+nU)^{2}(D+\epsilon _{0}+nU)^{2}\xi }}{T_{n}-D^{2}U\xi -U\xi (\epsilon
_{0}+nU)(2D-\epsilon _{0}-nU)}  \label{r8}
\end{equation}%
where%
\begin{equation}
T_{n}=(\epsilon _{0}+nU)(D+\epsilon _{0}+nU)^{2}-D^{2}U\xi  \label{r9}
\end{equation}%
\begin{figure}[tbp]
\centering\includegraphics[width=80mm,height=50mm]{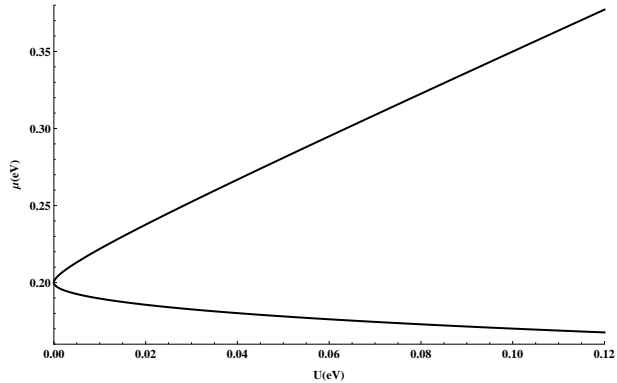}
\caption{Theoretical magnetic boundaries for adatom adsorbed in top site for 
$\protect\epsilon _{0}/D=0.029$ and $V/D=0.14$. }
\end{figure}
where the limiting cases must be carried out $n=0$ and $n=1$ and the $+$
solution for both cases. The solutions are shown as a function of $U$ for $%
\epsilon _{0}/D=0.029$ and $V/D=0.14$ in the top site adsorption. From the
figure, the magnetic boundaries coincide for large $U$ but disagree for $%
U\rightarrow 0$. The reason of this is that $n_{1}=n_{2}$ is the trivial
solution with $U=0$, then this point should appears in the magnetic
boundary. In turn, the magnetic boundary occurs for a critical $n_{c}$,
which, from the figure \ref{mag1}, is not exactly $1$ or $0$, then the
obtained \ critical curves enlarge the range of $\mu $ for fixed $U$.

The high sensitivity of the induced magnetism in the impurity for small $U$
values and small range of $\mu $ implies that spin polarization can be
induced by tuning the Fermi energy. One possible way to achieve this is by
field effect gating \cite{graph}. Using 300nm SiO$_{2}$ as dielectric
material, a gate potential can be applied between the sample of graphene
with adsorbed magnetic adatoms and the gate electrode (highly doped Si).
This applied gate voltage can shift the Fermi level and creates an
electrostactic potential between the sample. It is well known that maximum
voltage drop occurs across the SiO$_{2}$ and the conversion factor from the
gate potential to $\mu $ is very slow, $\sim 0.003$, which implies that in
order to change $\mu $ by $300$meV, a gate potential of $100$ V is needed.
Another way to increase the conversion factor is by using electrolyte gating
(see \cite{lu1}, \cite{das}), which allows to obtain a Fermi energy $\mu
=\hbar v_{F}\sqrt{\pi n}$, where $v_{F}$ is the Fermi velocity of graphene
and $n$ is the electron concentration and a gate potential $V_{G}=\mu /e+%
\frac{ne}{C_{TG}}$, where $e$ is the electron charge and $C_{TG}$ is the
geometrical capacitance which can be approximated as $C_{TG}=2.2\times
10^{-6}$F cm$^{-2}$ \cite{graph}. Transition elements and molecules that
usually do not magnetize when introduced in ordinary metals can actually
become magnetic in graphene (\cite{duffy}, \cite{lee}). In turn, an
enhancement of the local moment is harder for adatoms with a very large $U$
and which show a large local moment when hybridized with metals \cite{man},
but may be easily achieved in adatoms which are not usually magnetic and
exhibit a local moment in graphene \cite{intro1} and this magnetic moment
can be tuned by applying a gate voltage in order to use in spintronic
devices.

\section{Conclusions}

In this work, we have examined the conditions under which a transition metal
adatom adsorbed in top and hollow sites on graphene can form a local
magnetic moment. We find that due to the anomalous broadening of the adatom
local electronic states, moment formation is much easier in graphene. In
turn, for hollow site adsorption, magnetization appears for negligible $%
U\sim 0.002$eV. Theoretical curves for the magnetic boundaries in $\mu $-$U$
diagram are obtained, showing a wide range of possible values of $\mu $ for
fixed $U$ at which magnetization can be achieved. In turn, this magnetic
moment can be controlled by a field effect gating for the use in spintronics.

\section{Acknowledgment}

This paper was partially supported by grants of CONICET (Argentina National
Research Council) and Universidad Nacional del Sur (UNS) and by ANPCyT
through PICT 1770, and PIP-CONICET Nos. 114-200901-00272 and
114-200901-00068 research grants, as well as by SGCyT-UNS., J. S. A. and L.
S. are members of CONICET., F. E. is a fellow researcher at this institution.

\section{Author contributions}

All authors contributed equally to all aspects of this work.

\section{Appendix}

In order to compute the functions of eq.(\ref{6.05}), by using eq.(\ref{6.02}%
) we can write

\begin{equation}
\left\vert \Theta _{\mathbf{k\alpha }}\right\vert ^{2}=\frac{V^{2}}{2}\left(
2\left\vert \phi _{\mathbf{k}}\right\vert ^{2}+\frac{\alpha (-1)^{\gamma }}{%
\left\vert \phi _{\mathbf{k}}\right\vert }(\phi _{\mathbf{k}}^{3}+\phi _{%
\mathbf{k}}^{\ast 3})\right)  \label{ap1}
\end{equation}%
by writing%
\begin{equation}
\Delta _{H}=\overset{}{\underset{\alpha =\pm 1,\mathbf{k}}{\sum }}\frac{%
\left\vert \Theta _{\mathbf{k\alpha }}\right\vert ^{2}}{\omega -\alpha
\epsilon _{\mathbf{k}}}=V^{2}\overset{}{\underset{\mathbf{k}}{\sum }}\frac{%
2\left\vert \phi _{\mathbf{k}}\right\vert ^{2}\omega +t(-1)^{\gamma }(\phi _{%
\mathbf{k}}^{3}+\phi _{\mathbf{k}}^{\ast 3})}{\omega ^{2}-\epsilon _{\mathbf{%
k}}^{2}}  \label{ap1.1}
\end{equation}%
and by expanding $\phi _{\mathbf{k}}^{3}+\phi _{\mathbf{k}}^{\ast 3}$ around
the $K$ point, $\phi _{\mathbf{k}}^{3}+\phi _{\mathbf{k}}^{\ast 3}\sim \frac{%
6v_{F}^{3}}{t^{3}}k_{x}^{2}k_{y}+O(k^{4}),$ then by expanding the numerator
and denominator up to second order in $k$, $\Delta $ reads%
\begin{equation}
\Delta _{H}=\frac{2V^{2}\omega }{t^{2}}\overset{}{\underset{\mathbf{k}}{\sum 
}}\frac{v_{F}^{2}k^{2}}{\omega ^{2}-v_{F}^{2}k^{2}}=V^{2}\left[ a(\omega
)-ib(\omega )\right]  \label{ap2}
\end{equation}%
where%
\begin{equation}
a(\omega )=\frac{\omega }{t^{2}D^{2}}\left[ D^{2}+\omega ^{2}\ln (\frac{%
\omega ^{2}}{\omega ^{2}-D^{2}})\right]  \label{ap3}
\end{equation}%
and%
\begin{equation}
b(\omega )=\frac{\pi }{t^{2}D^{2}}\left\vert \omega \right\vert ^{3}\theta
(D-\left\vert \omega \right\vert )  \label{ap4}
\end{equation}%
This result should be compared with the hibridization for an adatom adsorbed
in a top site (see \cite{ucho2}). It should be noted that the above
approximation do not depends with the orbital index $\gamma $.

\bigskip

\bigskip

\bigskip

\bigskip

\end{document}